\documentclass[apjl, a4]{emulateapj}
\usepackage{natbib}
\usepackage{xspace}
\usepackage{graphicx}
\usepackage{hyperref}
\usepackage[normalem]{ulem}

\usepackage{xcolor}
\hypersetup{
    colorlinks,
    linkcolor={red!50!red},
    citecolor={blue!50!blue},
    urlcolor={blue!80!black}
}

\def\del#1{{}}

\newcommand{\e}{e}
\newcommand{\p}{p}
\newcommand{\CR}{\mathrm{cr}}
\newcommand{\eps}{\varepsilon}
\newcommand{\IC}{\rmn{IC}}
\newcommand{\B}{{\mathcal B}}
\newcommand{\rmn}{\mathrm}
\newcommand{\msun}{\mathrm{M}_\odot}
\newcommand{\Fermi}{{\em Fermi}\xspace}

\shorttitle{Simulating Gamma-ray Emission in Star-forming Galaxies}
\shortauthors{Pfrommer et al.}

\begin{document}

\title{Simulating Gamma-ray Emission in Star-forming Galaxies}

\author{Christoph Pfrommer\altaffilmark{1,2}, R{\"u}diger Pakmor\altaffilmark{2}, Christine M.~Simpson\altaffilmark{2} and Volker Springel\altaffilmark{2,3}}
\email{cpfrommer@aip.de}

\altaffiltext{1}{Leibniz-Institut f{\"u}r Astrophysik Potsdam (AIP), An der Sternwarte 16, 14482 Potsdam, Germany}
\altaffiltext{2}{Heidelberg Institute for Theoretical Studies, Schloss-Wolfsbrunnenweg 35, 69118 Heidelberg, Germany}
\altaffiltext{3}{Zentrum f\"ur Astronomie der Universit\"at Heidelberg, ARI, M\"onchhofstr. 12-14, 69120 Heidelberg, Germany}

\begin{abstract}
Star forming galaxies emit GeV- and TeV-gamma rays that are thought to originate
from hadronic interactions of cosmic-ray (CR) nuclei with the interstellar
medium. To understand the emission, we have used the moving mesh code {\sc
  Arepo} to perform magneto-hydrodynamical galaxy formation simulations with
self-consistent CR physics. Our galaxy models exhibit a first burst of star
formation that injects CRs at supernovae. Once CRs have sufficiently accumulated
in our Milky-Way like galaxy, their buoyancy force overcomes the magnetic
tension of the toroidal disk field. As field lines open up, they enable
anisotropically diffusing CRs to escape into the halo and to accelerate a
bubble-like, CR-dominated outflow. However, these bubbles are invisible in our
simulated gamma-ray maps of hadronic pion-decay and secondary inverse-Compton
emission because of low gas density in the outflows. By adopting a
phenomenological relation between star formation rate (SFR) and far-infrared
emission and assuming that gamma rays mainly originate from decaying pions, our
simulated galaxies can reproduce the observed tight relation between
far-infrared and gamma-ray emission, independent of whether we account for
anisotropic CR diffusion. This demonstrates that uncertainties in modeling
active CR transport processes only play a minor role in predicting gamma-ray
emission from galaxies.  We find that in starbursts, most of the CR energy is
``calorimetrically'' lost to hadronic interactions. In contrast, the gamma-ray
emission deviates from this calorimetric property at low SFRs due to adiabatic
losses, which cannot be identified in traditional one-zone models.
\end{abstract}


\section{Introduction}
\label{sec:intro}

CRs, magnetic fields and turbulent motions contribute equally to the total
midplane pressure in the Milky~Way \citep{1990ApJ...365..544B}. This could
originate from a self-regulated feedback loop and may even suggest that CRs 
play an active role in shaping galaxies by altering the dynamics and driving
galactic winds through their gradient pressure force, as suggested by
theoretical works \citep{1991A+A...245...79B,2008ApJ...674..258E}
three-dimensional simulations of galaxies
\citep{2012MNRAS.423.2374U,2013ApJ...777L..38H,2013ApJ...777L..16B,
  2014MNRAS.437.3312S,Pakmor2016b} and the interstellar medium (ISM)
\citep{2016ApJ...816L..19G,Simpson2016}. This idea can be scrutinized by
studying CR-induced radiative processes with the goal to extract the CR pressure
from the non-thermal emission at radio and gamma-ray energies.

Most of the galactic diffuse emission is generated by massive stars;
preferentially in the beginning and at the end of their lives. Young massive
stars emit mostly ultraviolet (UV) photons, which are absorbed by dust
grains. This radiation is then re-emitted in the far infrared (FIR) provided the
dust is optically thick to UV photons. This is the case in actively star-forming
galaxies, supporting the phenomenological correlation of the FIR emission with
the SFR \citep{1998ARA+A..36..189K}. After their fuel is exhausted, massive
stars explode as core-collapse supernovae, whose remnants are believed to
accelerate CR protons and electrons in galaxies through diffusive shock
acceleration.

CR electrons gyrate in the interstellar magnetic field and emit radio
synchrotron radiation that closely correlates with the total FIR luminosity of
galaxies over five orders of magnitude in luminosity
\citep{2003ApJ...586..794B}. The same radio-emitting CR electrons can generate
high-energy gamma-ray emission either through free-free transitions in the
vicinity of gas nuclei (bremsstrahlung) or by inverse Compton (IC) scattering
off of radiation fields. Hadronic collisions between CR nuclei and
the ISM also produce gamma rays alongside other decay products via pion decay:
\begin{eqnarray}
  \pi^0&\rightarrow&2\gamma\,,\nonumber\\  
  \pi^\pm&\rightarrow&\mu^\pm+\nu_{\mu}/\bar{\nu}_{\mu}\rightarrow\,e^\pm+\nu_{e}/\bar{\nu}_{e}+\nu_{\mu}+\bar{\nu}_{\mu}\,.\nonumber
\end{eqnarray}
These secondary relativistic electrons and positrons (hereafter simply
electrons) can emit radio-synchrotron emission in the presence of ubiquitous
galactic magnetic fields as well as up-scatter ambient photon fields into the
gamma-ray regime through IC interactions.

Recently, nearby starburst galaxies have been detected in high-energy gamma rays
by space-based \citep{2010ApJ...709L.152A} and imaging air-Cerenkov telescopes
\citep{2009Natur.462..770V,2009Sci...326.1080A}. Upper limits and detections of
the galactic gamma-ray emission by the \Fermi telescope revealed a tight
FIR--gamma-ray relation \citep{2012ApJ...755..164A, 2016MNRAS.463.1068R}. The
gamma-ray emission enables testing whether starbursts act as proton
``calorimeters'' \citep[where most of the CR energy is lost to hadronic
  interactions,][]{2011ApJ...734..107L, 2016arXiv161207290W} and whether CRs are
dynamically important in starbursts \citep{2016MNRAS.457L..29Y}.

Here, we present high-resolution MHD simulations with CR physics of forming disk
galaxies and compute the gamma-ray emission. This enables us to critically
assess simplifying assumptions made in previous one-zone analyses. We describe
our simulations in \S\ref{sec:simulations} and lay out our methodology to
compute pion-decay and IC emission in \S\ref{sec:radiation}.  We analyze
gamma-ray emission maps and the FIR--gamma-ray relation in \S\ref{sec:results}
and conclude in \S\ref{sec:conclusion}.

\section{Simulations}
\label{sec:simulations}

We simulate the formation of isolated disks in dark-matter (DM) halos that range
in mass from $10^{10}$ to $10^{12}~\msun$. We use the second-order accurate,
adaptive moving-mesh code \textsc{Arepo} \citep{2010MNRAS.401..791S,
  2016MNRAS.455.1134P}.  We model the ISM with an effective equation of state
with radiative cooling and star formation using a probabilistic approach
\citep{2003MNRAS.339..289S}. Magnetic fields are treated with ideal MHD
\citep{2013MNRAS.432..176P}.

CRs are modelled as a relativistic fluid with a constant adiabatic index of
$4/3$ in a two-fluid approximation \citep{2017MNRAS.465.4500P}.  We describe CR
generation at remnants of core-collapse supernovae by instantaneously injecting
all CR energy produced by a stellar particle’s population through this channel
with an energy efficiency of $\zeta_{\rmn{SN}}=0.1$ into its surroundings
immediately after birth.  We account for adiabatic changes of the CR energy as
well as CR cooling via Coulomb and hadronic CR interactions, assuming an
equilibrium momentum distribution \citep{2017MNRAS.465.4500P}. To bracket the
uncertainty regarding CR transport, we simulate two models. In model {\em
  CR~adv}, we only advect CRs with the gas while model {\em CR~diff}
additionally accounts for anisotropic diffusion relative to the rest frame of
the gas with a diffusion coefficient of $10^{28}\mathrm{cm^{2}\,s^{-1}}$ along
the magnetic field and no diffusion perpendicular to it \citep{Pakmor2016a}. Our
simulations do not include CR streaming \citep[unlike][]{2017MNRAS.467..906W}.

DM halos are modelled as static NFW \citep{1997ApJ...490..493N} profiles with a
fixed concentration parameter of $c_{200}=12$ across our halo mass range and one
$10^{12}~\msun$ halo with $c_{200}=7$. The gas is initially in hydrostatic
equilibrium with the DM potential. The halo carries a small amount of angular
momentum, parametrized by the spin parameter $\lambda=0.05$.  In all cases, we
adopt a baryon mass fraction of $\Omega_\rmn{b}/\Omega_\rmn{m}=0.155$.  We
assume a uniform, homogeneous seed field along the $x$-axis with strength
$10^{-12}~\mathrm{G}$ and no CRs in the initial conditions.

We start our simulations with $10^7$ gas cells inside the virial radius, each of
which has a target mass of $155~\msun\times\,M_{10}$, where
$M_{10}=M_{200}/(10^{10}~\msun)$.  This target gas mass corresponds to the
typical mass of a stellar population particle. We require that the mass of all
cells is within a factor of two of the target mass by explicitly refining and
de-refining cells. We additionally require adjacent cells to differ in volume by
less than a factor of $10$ and refine the larger cell if this condition is
violated.

\section{Radiative Processes}
\label{sec:radiation}

\subsection{Pion-decay gamma rays}

Pion-decay gamma rays arise from inelastic CR interactions with thermal nuclei.
We assume that the one-dimensional CR momentum spectrum per volume element,
$d^3x$, follows a power-law:
\begin{equation}
  \label{eq:fp}
  \frac{d^4N}{dp\,d^3x}\equiv{f}_\p(p)\equiv4{\pi}p^2f_\p^{(3)}(p)=C_{\p}p^{-\alpha}\,\theta(p-q),
\end{equation}
where $p=P_\p/(m_{\p}c)$ is the dimensionless proton momentum, $\theta(a)$
denotes the Heaviside step function, $q$ is the low-momentum cutoff and $\alpha$
is the CR spectral index. The normalization is determined by solving the CR
energy density ($\eps_\CR$) for $C_\p$:
\begin{eqnarray}
\label{eq:eps}
\lefteqn{\eps_\CR=\int_0^{\infty}f_\p(p)\,E(p)dp=\frac{C_\p\,m_{\p}c^2}{\alpha-1}}\\
&&\,\times\left[\frac{1}{2}\,\B_\frac{1}{1+q^2}\left(\frac{\alpha-2}{2},\frac{3-\alpha}{2}\right)+
q^{\alpha-1}\left(\sqrt{1+q^2}-1\right)\right],\nonumber
\end{eqnarray}
where $E(p)=(\sqrt{1+p^2}-1)\,m_{\p}c^2$ is the kinetic proton energy and
$\B_y(a,b)$ denotes the incomplete beta function \citep{1965hmfw.book.....A},
assuming $\alpha>2$.

The omnidirectional gamma-ray source function from decaying pions
($s_{\pi^0-\gamma}$) is \citep{2004A+A...413...17P}:
\begin{eqnarray}
\label{eq:s_gamma}
\lefteqn{\!\!\!\!\!\!\!\!\!\!
  \frac{d^5N_\gamma}{dE_{\gamma}\,dt\,d^3x}\equiv{s}_{\pi^0-\gamma}(E_\gamma)\simeq
\frac{2^4C_\p}{3 \alpha}\,\frac{\sigma_{pp}\,n_{n}}{m_{\p}c}\,
\left(\frac{m_\p}{2m_{\pi^0}}\right)^{\alpha}}\nonumber\\
  &&\quad\times\left[\left(\frac{2E_\gamma}{m_{\pi^0}\,c^2}\right)^{\delta}+
      \left(\frac{2E_\gamma}{m_{\pi^0}\,c^2}\right)^{-\delta}\right]^{-\alpha/\delta},
\end{eqnarray}
where $m_{\pi^0}$ is the pion rest mass and
$n_{n}=n_\rmn{H}+4n_\rmn{He}=\rho/m_\p$ is the target density of nucleons,
neglecting metals.  This treatment accounts for all physical processes at the
pion production threshold, which are parametrized by the shape parameter $\delta$
and the effective cross section $\sigma_{pp}$ in terms of the photon index
$\alpha_\gamma=\alpha$ according to
\begin{eqnarray}
\label{eq:delta}
\delta&\simeq&0.14\,\alpha^{-1.6}+0.44\qquad\mbox{and}\\
\label{eq:sigmapp}
\sigma_{pp}&\simeq&32\times\left(0.96+\rmn{e}^{4.4\,-\,2.4\,\alpha}\right)~\mbox{mbarn}. 
\end{eqnarray}

The energy-integrated gamma-ray emissivity for the energy band $[E_1,E_2]$ (in
units of erg~s$^{-1}$~cm$^{-3}$) that results from pion decay is given by
\begin{eqnarray}
\label{eq:Lambda_gamma}
\lefteqn{\Lambda_{\pi^0-\gamma}(E_1,E_2)=\int_{E_1}^{E_2}s_{\pi^0-\gamma}(E_\gamma)E_\gamma\,dE_\gamma}\\
\label{eq:Lambda_gamma2}
&&\!\!=\frac{2C_\p}{3\alpha\delta}
\frac{m_{\pi^0}^2c^3\sigma_{pp}n_{n}}{m_\p}\left(\frac{m_\p}{2m_{\pi^0}}\right)^{\alpha}
\left[\mathcal{B}_y\left(\frac{\alpha+2}{2\delta},\frac{\alpha-2}{2\delta}\right)\right]_{y_1}^{y_2},\nonumber\\
&&y_i=\left[1+\left(\frac{m_{\pi^0}c^2}{2E_i}\right)^{2\delta}\right]^{-1}\mbox{~for~}i\in\{1,2\}.
\end{eqnarray}

\subsection{Secondary inverse Compton emission}

The mean energies of isotropically scattered IC photons and scattering CR electrons
are related by
\begin{equation}
  \label{eq:ICphoton}
  h\nu_\rmn{IC}=\frac{4}{3}\,h\nu_\rmn{init}\,\gamma^2\simeq 
  1~\mbox{GeV}\,\frac{\nu_\rmn{init}}{\nu_\rmn{CMB}}\,\left(\frac{\gamma}{10^6}\right)^2,
\end{equation}
where the particle kinetic energy $E/(m_{\e}c^2)=\gamma-1$ is defined in terms
of the Lorentz factor $\gamma$. We adopt cosmic microwave background (CMB)
photons with a characteristic energy $h\nu_\rmn{CMB}\simeq0.66$~meV as the
source for IC emission using Wien's displacement law. The IC cooling time of
these relativistic electrons is $t_\rmn{IC}\sim2~\rmn{Myr}$ and thus shorter
than both the CR residency time in our Galaxy and the dynamical time scales in
the warm and coronal phases of the ISM. Hence, the steady-state approximation
for modeling secondary IC emission is justified.  Note that we neglect IC
interactions with starlight and FIR photons: including them would not change any
conclusions since the steady state IC emissivity does not depend on the photon
energy density (in the IC-dominated scattering regime), which solely depends on
the CR electron energy density.  Throughout the Letter, we neglect primary IC
emission, which has a subdominant contribution to the \Fermi-band luminosity
\citep[but may be important outside the disk,][]{2015A+A...581A.126S}.

At high momenta ($P_\e>\mbox{GeV}/c$), the injection of secondaries is balanced by
IC and synchrotron cooling, which results in an equilibrium distribution of
secondary CR electrons \citep{2008MNRAS.385.1211P},
\begin{eqnarray}
\label{eq:fe_hadr}
f_\e(p)\,dp&=&C_{\e}p^{-\alpha_\e}\,dp\\
\label{eq:C_e}
C_\e&=&\frac{16^{2-\alpha_\e}\sigma_{pp}\,n_{n} C_\p\,m_{\e}c^2}
     {(\alpha_\e-2)\,\sigma_\rmn{T}\,(\eps_B+\eps_\rmn{ph})}\left(\frac{m_\p}{m_\e}\right)^{\alpha_\e-2},
\end{eqnarray}
where we redefined $p=P_\e/(m_{\e}c)$ as the dimensionless electron momentum,
$\alpha_\e=\alpha+1$ is the electron spectral index, $\sigma_\rmn{T}$ is the
Thomson cross section, and $\eps_B$ and $\eps_\rmn{ph}$ are the energy densities
of the magnetic and photon fields, respectively.

The energy-integrated gamma-ray emissivity for an isotropic power law
distribution of CR electrons of equation~(\ref{eq:fe_hadr}) that
Compton-upscatters CMB photons is (derived from eq. (7.31) in
\citet{1979rpa..book.....R}, in the case of Thomson scattering),
\begin{eqnarray}
\label{eq:IC}
\!\!\!\!\!
\Lambda_{\IC-\gamma}(E_1,E_2)&=&\int_{E_1}^{E_2}s_{\IC-\gamma}(E_\gamma)E_\gamma\,dE_{\gamma}\\
&=&\Lambda_0\,f_\IC(\alpha_e)
\left[\left(\frac{E_\gamma}{kT_\mathrm{CMB}}\right)^{-\alpha_\nu}\right]_{E_2}^{E_1},\\
f_\mathrm{IC}(\alpha_e)&=&\frac{2^{\alpha_\e+3}\,(\alpha_\e^2+4\,\alpha_\e+11)}
  {(\alpha_\e+3)^2\,(\alpha_e+5)\,(\alpha_e+1)}\nonumber\\
&&\times\,\Gamma\left(\frac{\alpha_e+5}{2}\right)\,\zeta\left(\frac{\alpha_e+5}{2}\right),\\
  \mbox{and}~\Lambda_0&=&\frac{16\,\pi^2\,r_\e^2\,C_\e\,
    \left(kT_\mathrm{CMB}\right)^4\,}{(\alpha_\e-3)\,h^3\,c^2}, 
\end{eqnarray}\\
where $\alpha_\nu=(\alpha_\e-1)/2$ denotes the IC spectral index,
$T_{\rmn{CMB}}=2.725$~K is the CMB temperature at the present time,
$r_\e=e^2/(m_{\e}c^2)$ the classical electron radius, $e$ is the elementary
charge, $h$ is Planck's constant and $\Gamma(a)$ and $\zeta(a)$ are the gamma
and Riemann $\zeta$ functions, respectively \citep{1965hmfw.book.....A}.  The
gamma-ray luminosities follow as a result of volume integrations of the
emissivities, $L_{i-\gamma}=\int\Lambda_{i-\gamma}d^3x$ with
$i\in\{\rmn{\pi^0,\rmn{IC}}\}$.

\section{Results}
\label{sec:results}

\subsection{CR-driven outflows in Milky~Way-like galaxy}

\begin{figure*}
\centering
\includegraphics[width=0.99\textwidth]{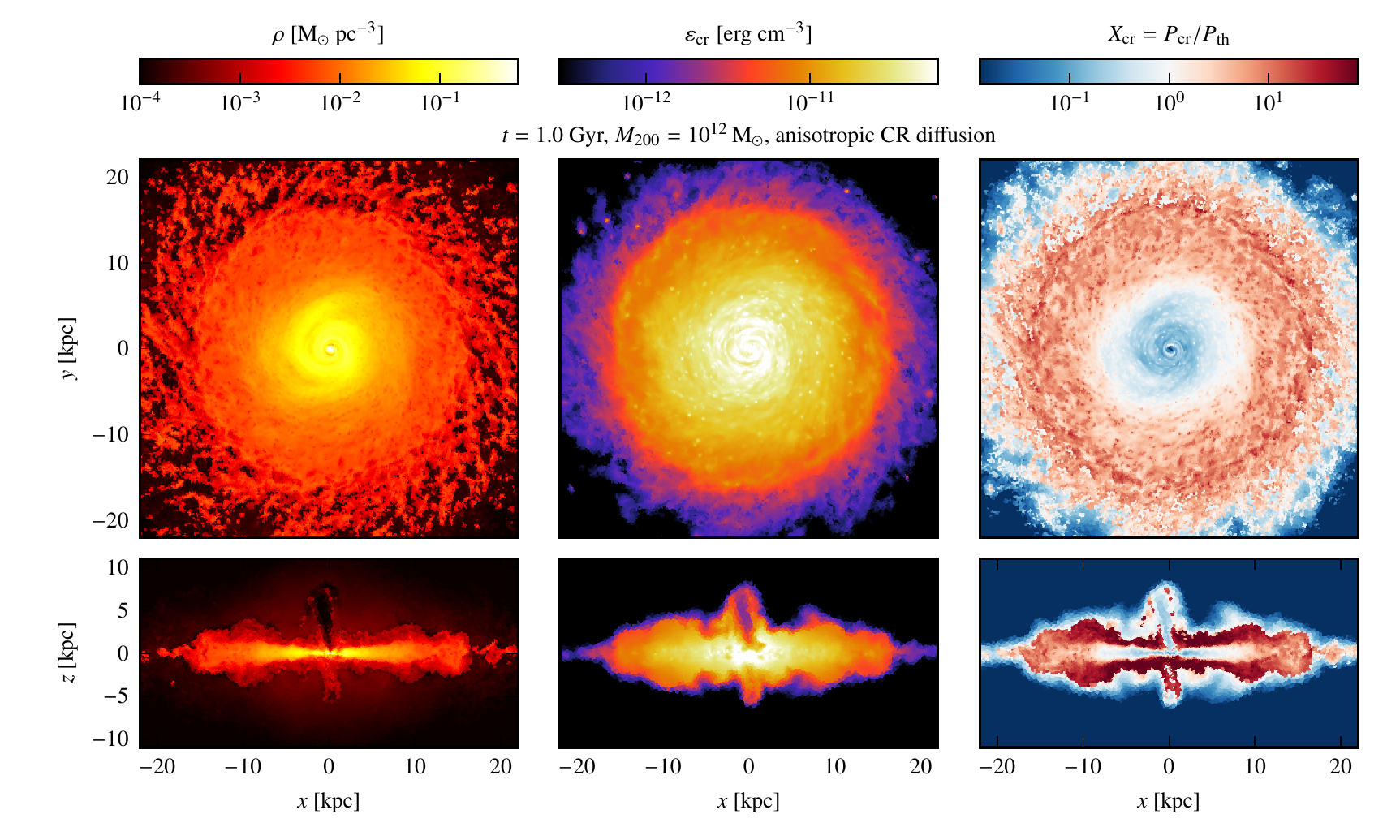}
\includegraphics[width=0.99\textwidth]{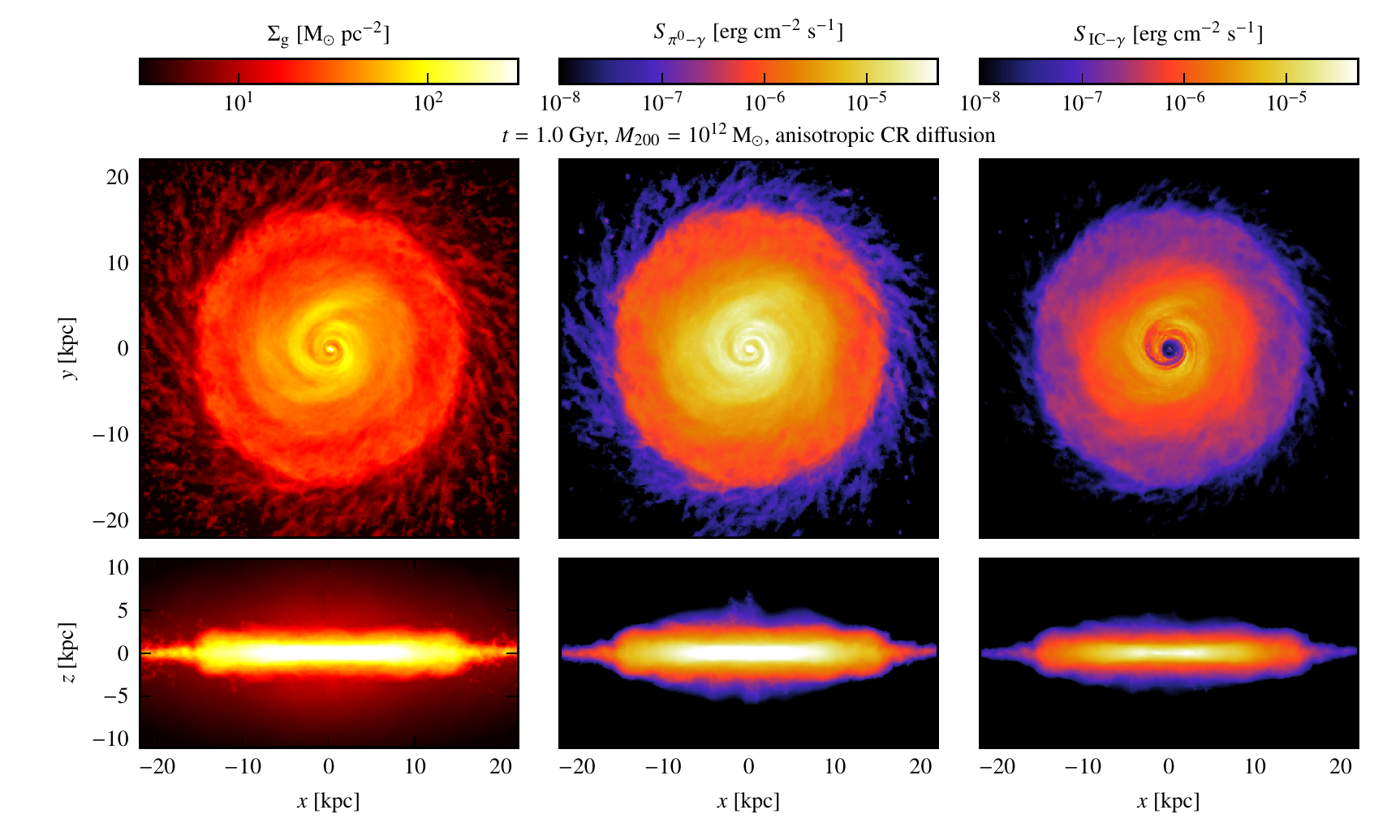}
\caption{Properties of the gas disk in our Milky~Way-like galaxy
  ($M_{200}=10^{12}~\msun$) at 1~Gyr, shortly after the onset of a CR-driven
  central outflow. The MHD simulations account for CR injection at supernova
  remnants and follow their advection with the gas and anisotropic diffusion
  along the magnetic fields relative to the gas.  In the top panels, we show
  cross-sections of gas properties in the mid-plane of the disk (face-on views)
  and vertical cut-planes through the centre (edge-on views) of the gas density
  (left), CR energy density (middle), and CR-to-thermal pressure ratio
  (right). In the bottom panels, we show face-on and edge-on projections of
  observables: gas surface mass density (left), pion decay-induced gamma-ray
  surface brightness from 0.1 to 100~GeV (middle) and secondary IC surface
  brightness in the same energy band (right). Both gamma-ray maps do not show
  the CR-loaded outflow in the form of hadronic gamma-ray bubbles.  }
\label{fig:maps}
\end{figure*}

We first analyze our simulation of the Milky~Way-like galaxy using the model
{\em CR~diff} ($M_{200}=10^{12}~\msun$, $c_{200}=7$). At the beginning of our
simulation, radiative cooling diminishes pressure support of the central dense
gas, starting collapse while conserving the gas' specific angular
momentum. After settling into a rotationally supported disk, gas is compressed
by self-gravity to sufficiently high densities for star formation. During the
first collapse phase, a turbulent dynamo quickly grows a small-scale magnetic
field, which is further amplified by a large-scale dynamo that preferentially
grows a toroidal field in the disk \citep{Pakmor2016b, 2017MNRAS.469.3185P}.

CRs are injected into the ambient ISM surrounding stellar macro-particles,
providing the gas with additional non-thermal pressure. As CRs are advected and
diffused, they collect in the disk and start to dominate the pressure,
especially in the outer midplane of the disk and everywhere at the disk-halo
interface (Fig.~\ref{fig:maps}). This result points to the importance of
modeling CRs in simulations of galaxy formation that aim to capture their
dynamical evolution.  While our SFRs significantly vary over the simulation, our
galaxy properties are not changing too quickly on the timescale over which the
CRs lose energy and hence can be considered a reasonable proxy for observed
systems.

However, the CR pressure falls short of the thermal pressure in the inner part
of the disk. This is caused by over-cooling of our steady-state CR fluid in the
densest regions since the simulations assume a softer CR spectral index
($\alpha=2.2$) than the analysis ($\alpha=2.05$).  This will be improved upon
with future simulations that dynamically follow the CR spectrum. Moreover, we
adopt a single-phase ISM rather than a realistic multi-phase ISM. The latter
would minimize CR losses since CRs spend more time in the lower-density warm and
coronal phases, which dominate the volume.

After 1~Gyr, the CR pressure has increased to the point where the buoyancy force
overcomes the magnetic tension of the toroidal magnetic field which bends and
opens. CRs diffuse ahead of the gas into the halo and accelerate the gas,
thereby driving a bubble-like outflow (Fig.~\ref{fig:maps}).  The total
pressure, composed of thermal, CR, and magnetic pressures, declines smoothly
outwards, which demonstrates that the bubble edges are contact discontinuities
and not shocks.

\begin{figure}[t]
\centering
\includegraphics[width=0.49\textwidth]{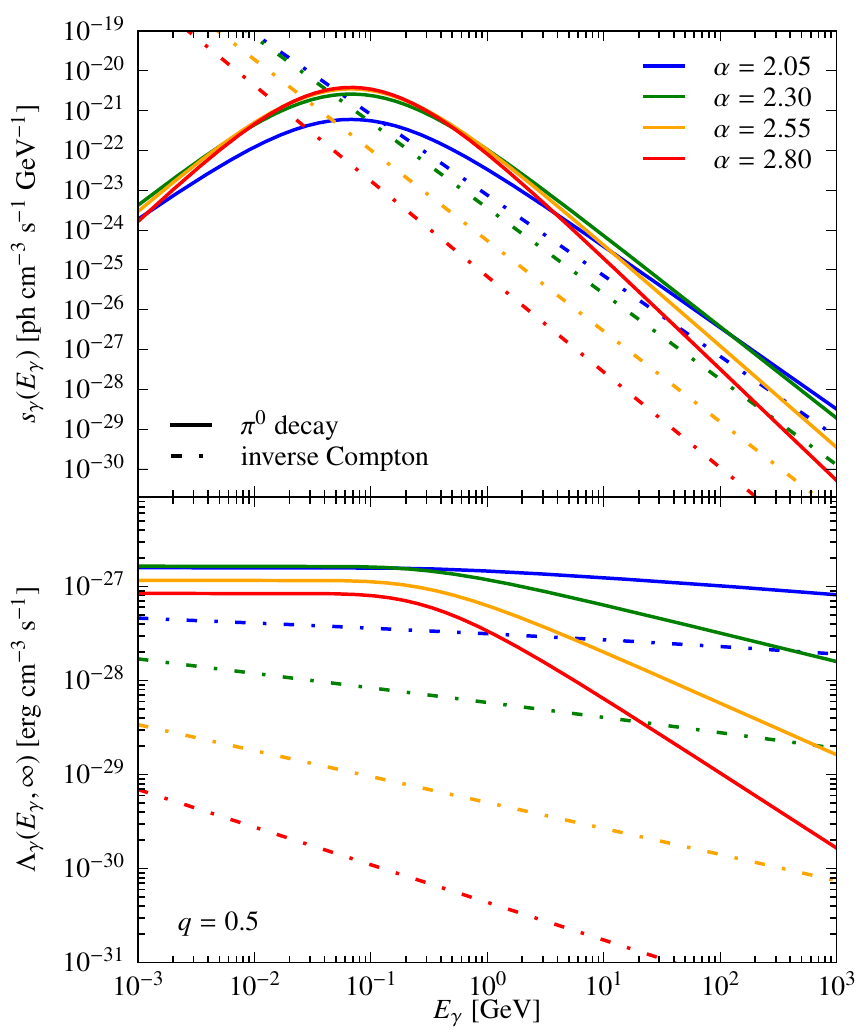}
\includegraphics[width=0.49\textwidth]{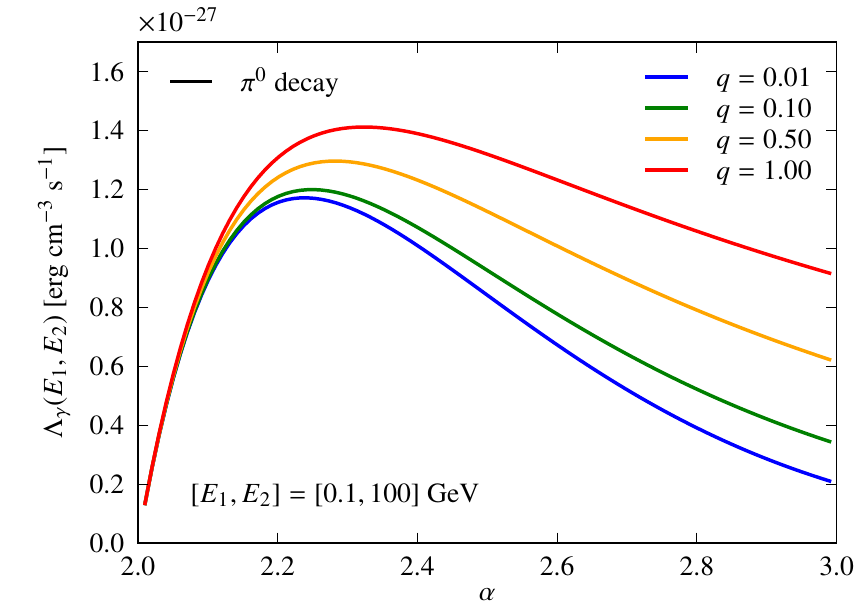}
\caption{Spectral distribution of the differential gamma-ray source function
  (top) and the energy-integrated gamma-ray emissivity, $\Lambda_\gamma$,
  vs.~energy (middle) that result from hadronic CR interactions for different CR
  spectral indices, $\alpha$ (colored differently). We also present
  $\Lambda_\gamma$ vs.~spectral index in the \Fermi-energy band (bottom) for
  various low-momentum cutoffs $q$ of the distribution function (colored
  differently). Shown are pion decay induced gamma-ray spectra (solid) and IC
  spectra from a steady-state secondary CR electron population (dash dotted) in
  the weak-field regime ($B=0$).  The model calculations assume a fixed CR
  energy density of $\eps_\CR=10^{-11}\,\rmn{erg~cm}^{-3}$ and target nucleon
  density $n_n=1\,\rmn{cm}^{-3}$, which both impact the absolute normalization
  of the spectra in equal measures. }
\label{fig:gamma_spectrum}
\end{figure}

We project the pion-decay and secondary IC emissivities along the
line-of-sight into face-on and edge-on views (Fig.~\ref{fig:maps}) according to
$S_{i-\gamma}(\mathbf{\hat{n}})=\int\Lambda_{i-\gamma}(r,\mathbf{\hat{n}})dr$
where $i\in\{\rmn{\pi^0,\rmn{IC}}\}$ and $\mathbf{\hat{n}}$ is a unit vector
perpendicular to the line-of-sight. Despite the high CR load of the outflow,
neither pion-decay nor IC gamma-ray emission show signatures of hadronic
gamma-ray bubbles because the fast outflow
($\upsilon_z\lesssim150~\rmn{km~s}^{-1}$) evacuates the gas in the bubble
region. This is consistent with findings by Bayesian non-parametric
reconstructions of the \Fermi sky which suggests that the \Fermi bubbles are of
leptonic origin \citep{2015A+A...581A.126S}.

Figure~\ref{fig:gamma_spectrum} compares the pion decay-induced gamma-ray
spectra to the IC spectra from an equilibrium secondary CR electron population
in the weak-field regime ($B=0$). This choice maximizes the IC yield of the
cooling electron population and keeps the pion decay-to-IC photon ratio at a
constant value that solely depends on the spectral index. Increasing the field
strength above the CMB equivalent field strength of $B_{\rmn{CMB}}=3.24\,\mu$G,
causes the electrons to cool preferentially by emitting radio synchrotron
radiation. Less energy is then emitted via IC interactions. This happens in our
Milky~Way-like galaxy for radii $r\lesssim3$~kpc at $t=1$~Gyr and the field
strength increases with time to tens of $\mu$G in the center due to the ongoing
magnetic dynamo. At larger radii in the low-$B$ regime,
$\Lambda_{\pi^0-\gamma}/\Lambda_{\rmn{IC}-\gamma}=4.3$ for $\alpha=2.05$. Due to
the strong IC suppression in the center, the pion-decay luminosity
($3.2\times10^{40}\,\rmn{erg~s}^{-1}$) dominates over the secondary IC
luminosity ($5.6\times10^{39}\,\rmn{erg~s}^{-1}$) by a factor of 5.7. Hence, we
neglect the IC contribution to the gamma-ray luminosity in what follows.

Throughout this Letter, we adopt $\alpha=2.05$ and $q=0.5$ which yields an
average value of $\Lambda_{\pi^0-\gamma}$ that is within a factor of 2.5 from
the extreme outliers for a broad variety of physically motivated values for
$\alpha$ and $q$ (see bottom panel of Fig.~\ref{fig:gamma_spectrum}). The
observed photon index is expected to be somewhat steeper \citep[matching the
  observed range of $\alpha_\gamma=2.1$ to 2.4,][]{2012ApJ...755..164A} due to a
combination of energy-dependent CR streaming and diffusion in Kolmogorov
turbulence.

\subsection{FIR--gamma-ray relation}

We now analyze our entire galaxy sample for both models {\em CR~adv} and {\em
  CR~diff} while fixing $c_{200}=12$ to ensure self-similarly evolving galaxies.
We believe that each of our galaxy simulations are good analogues of observed
galaxies as they go through different evolutionary phases: initial gas collapse
is immediately followed by a starburst that transitions to an intense
star-forming thick disk, which eventually settles to a quiescently star-forming
thin-disk galaxy.

After 100 Myr, the $10^{12}~\msun$ halo enters a starburst phase with a central
density of $3~\msun~\rmn{pc}^{-3}$ (averaged within a radius of
300~pc). Line-of-sight integration over this central region yields a central
surface mass density of $1.8\times10^3~\msun~\rmn{pc}^{-2}$ or
$0.4~\rmn{g~cm}^{-2}$, amounting to 1.6 (2.6) times the surface mass density
inferred from M~82 (NGC~253), see \citet[][]{2011ApJ...734..107L}. The
corresponding simulated SFR density is
$\approx40~\msun~\rmn{yr}^{-1}~\rmn{kpc}^{-3}$, which integrates to a global SFR
of $\approx120~\msun~\rmn{yr}^{-1}$, substantially larger than the SFRs in M~82
and NGC~253. The simulations reach a peak resolution (minimum cell radius) of
$\approx5$~pc, sufficient to resolve the central starburst region.

\begin{figure}[t!]
\centering
\includegraphics[width=0.49\textwidth]{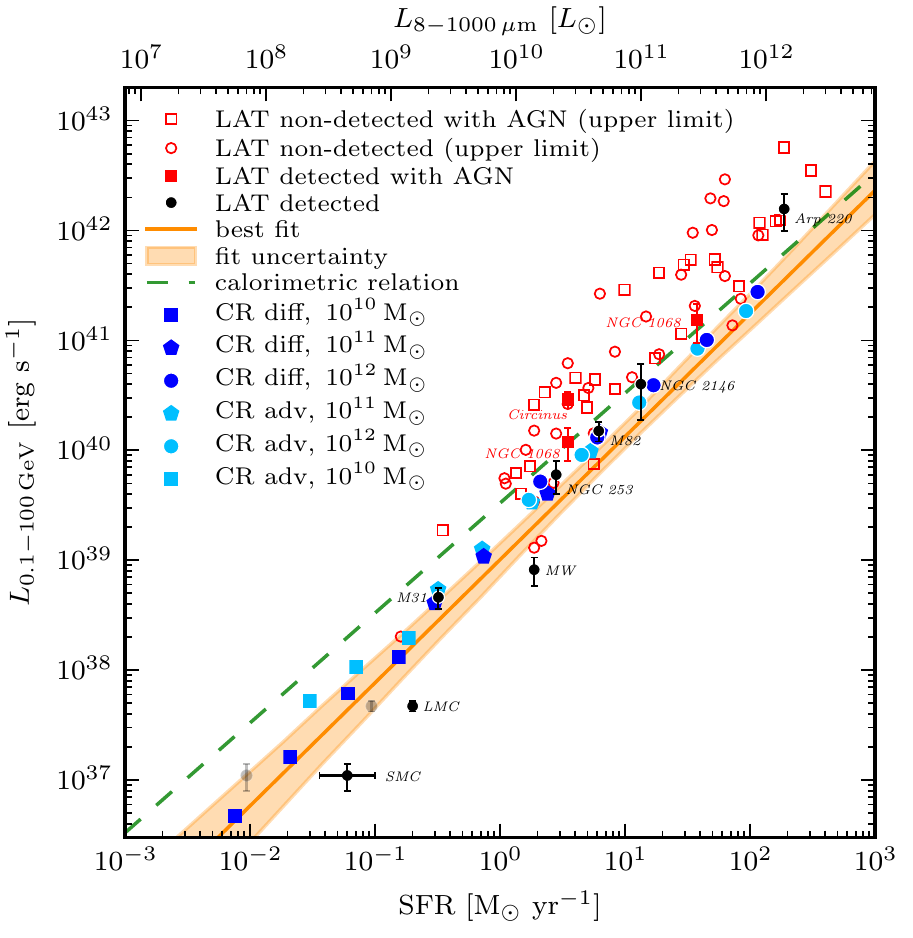}
\caption{Correlation between the gamma-ray luminosity ($L_{0.1-100\,\rmn{GeV}}$)
  and the SFR, respectively the FIR luminosity ($L_{8-1000\,\mu\rmn{m}}$) of
  star forming galaxies. Upper limits on the observable gamma-ray emission by
  \Fermi-LAT \citep[open red symbols,][]{2016MNRAS.463.1068R} are contrasted to
  gamma-ray detections from star-forming galaxies only (solid black) and with
  AGN emission (filled red); data are taken from \citet{2012ApJ...755..164A},
  except for NGC~2146 \citep{2014ApJ...794...26T} and Arp~220
  \citep{2016ApJ...823L..17G,2016ApJ...821L..20P}. See text for details on the
  FIR-to-SFR conversion.  We overplot the emission of our simulated galaxies
  that only account for advective CR transport (light blue) and simulations in
  which we additionally follow anisotropic CR diffusion (blue). Different
  symbols indicate simulations of differently sized galaxy halos. In each
  simulation, we sample the SFR in logarithmic steps of {e}. Note that our
  simulations fall on the best-fit observational FIR-gamma-ray correlation
  (orange) and start to deviate from the calorimetric relation (dashed green)
  for small SFRs.}
\label{fig:IRgamma}
\end{figure}

The total FIR luminosity ($8-1000~\mu$m) is a well-established tracer of the SFR
of spiral galaxies \citep{1998ApJ...498..541K} with a conversion rate
\citep{1998ARA+A..36..189K}
\begin{equation}
  \label{eq:FIR-SFR}
  \frac{\rmn{SFR}}{\msun~\rmn{yr}^{-1}}=\epsilon\,1.7\times10^{-10}\,\frac{L_{8-1000\,\mu\rmn{m}}}{L_\odot}.
\end{equation}
This assumes that thermal dust emission is a calorimetric measure of the
radiation of young stars, and the factor $\epsilon=0.79$ derives from the
\citet{2003ApJ...586L.133C} initial mass function
\citep[IMF,][]{2010MNRAS.407.1403C}.  While this conversion is reliable at
$L_{8-1000\,\mu\rmn{m}}>10^9~L_\odot$, it becomes progressively worse at smaller
FIR luminosities due to the lower metallicity and dust content, which implies a
low optical depth to IR photons and invalidates the calorimetric assumption
\citep{2003ApJ...586..794B}. Blindly applying the conversion yields the grey
data points in Fig.~\ref{fig:IRgamma} for the Small and Large~Magellanic~Clouds
(SMC, LMC). More reliable SFR estimates for the SMC range from
$0.036~\msun~\rmn{yr}^{-1}$ \citep[combining H$_\alpha$ and FIR emission,
  assuming a Chabrier~IMF,][]{2004A&A...414...69W} to $0.1~\msun~\rmn{yr}^{-1}$
\citep[UVBI photometry,][]{2004AJ....127.1531H} and yield
$0.2~\msun~\rmn{yr}^{-1}$ for the LMC \citep[UVBI
  photometry,][]{2009AJ....138.1243H}.

In Fig.~\ref{fig:IRgamma} we correlate the gamma-ray luminosity in the
\Fermi~band ($0.1-100$~GeV) to the SFR for all simulated galaxies at various
times (see Table~\ref{tab:Lgamma} and relate them to the FIR luminosity via
equation~\ref{eq:FIR-SFR}). We find very similar gamma-ray luminosities in our
models {\em CR~adv} and {\em CR~diff}, which both match the observed relation
$L_\gamma/(\rmn{erg\,s}^{-1})=8.9\times10^{27}(L_{8-100~\mu\rmn{m}}/L_\odot)^{1.12}$
\citep{2016MNRAS.463.1068R} for $L_{8-1000\,\mu\rmn{m}}>10^9~L_\odot$. The
simulations appear to overpredict the gamma-ray luminosity at the lowest
SFRs. The match may be improved by lowering the gas density in these halos or by
adopting a realistic multi-phase ISM, which becomes more porous towards lower
SFRs.

The relative contribution of hadronic losses to the total CR loss rate,
$\Gamma_{\rmn{hadr}}/\Gamma_{\rmn{tot}}$, sets the normalization of the
calorimetric relation, $L_\gamma\propto{L}_\CR\propto\rmn{SFR}$ ($L_\CR$ is the
CR luminosity).  Star-forming galaxies with
$\rmn{SFR}\gtrsim10~\msun~\rmn{yr}^{-1}$ are close to the calorimetric relation
(see Fig.~\ref{fig:IRgamma}). Instead, galaxies with lower SFRs fall below this
relation, indicating that non-hadronic CR energy losses start to become
relevant.

\begin{figure*}
\centering
\includegraphics[width=0.49\textwidth]{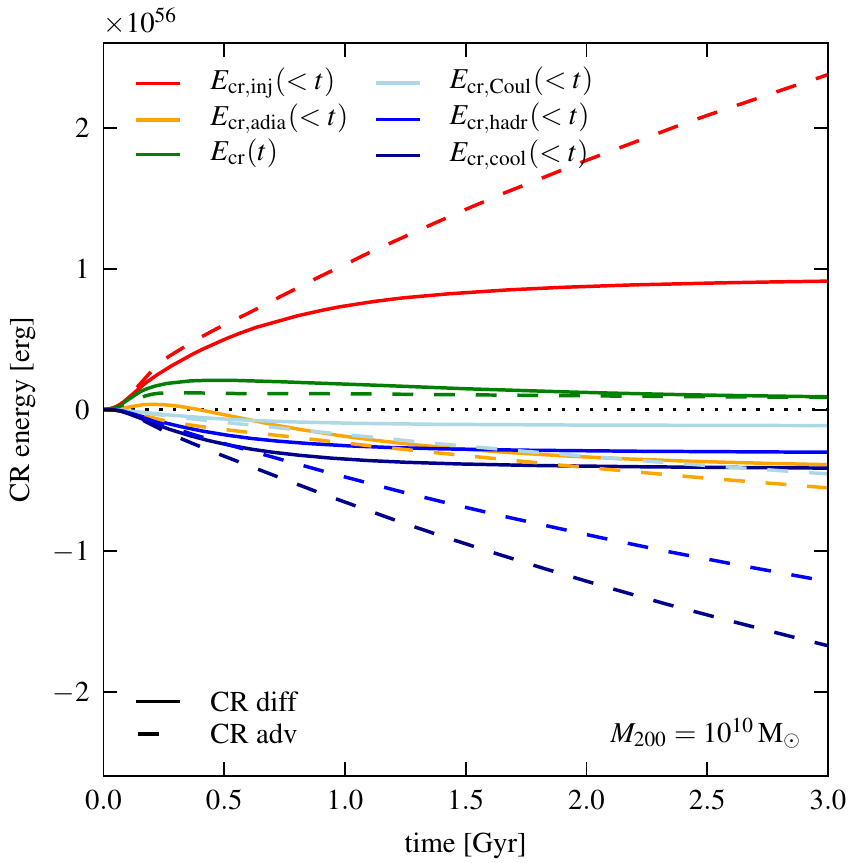}
\includegraphics[width=0.49\textwidth]{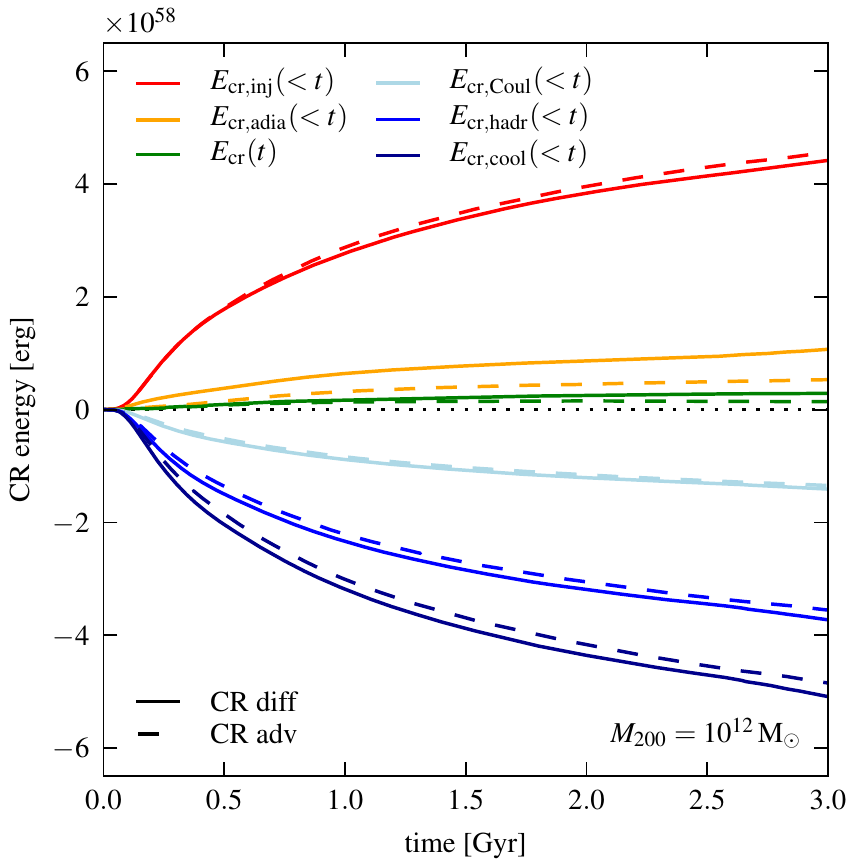}
\caption{Time evolution of different CR energy gain and loss processes for a
  dwarf galaxy of mass $10^{10}~\msun$ (left) and a Milky~Way-like galaxy of
  mass $10^{12}~\msun$ (right). We contrast our model of advective CR transport
  (dashed) to the model in which we additionally follow anisotropic CR diffusion
  (solid). While CRs experience a modest adiabatic energy gain in the Milky
  Way-like galaxy, they suffer a substantial adiabatic loss in the dwarf galaxy
  which even matches the non-adiabatic CR loss in model {\em CR~diff}. This
  causes the FIR--gamma-ray relation to deviate from the calorimetric relation.}
\label{fig:crenergy}
\end{figure*}

We explore different CR energy gain and loss processes in
Fig.~\ref{fig:crenergy}. In the Milky~Way-like halo, the non-adiabatic CR loss
processes (hadronic and Coulomb interactions) cool most of the CR energy that is
injected; with cooling rates of
$\Gamma_{\rmn{hadr}}=7.44\times10^{-16}~\rmn{s}^{-1}$ and
$\Gamma_{\rmn{Coul}}=2.78\times10^{-16}~\rmn{s}^{-1}$, respectively
\citep{2017MNRAS.465.4500P}. The (subdominant) adiabatic gains in the {\em
  CR~diff} model are twice those of the {\em CR~adv} model as CRs that are
diffusing from the dense ISM into the halo are caught by the accretion flow and
get advected back onto the disk. Hence, the calorimetric assumption for these
high-mass galaxies is justified.

This is in stark contrast to the dwarf galaxy simulation ($10^{10}~\msun$ halo),
where adiabatic and non-adiabatic losses are equally strong in the {\em CR~diff}
model. This is the main reason why these slowly simmering star-forming galaxies
deviate from the calorimetric relation. In our dwarf-galaxy simulation in the
{\em CR~adv} model, the SFR levels off after 0.3~Gyr. This causes an almost
linear increase of the CR injection rate that is counteracted by the increasing
non-adiabatic CR cooling rates. Those dominate over the adiabatic CR losses at
late times and thus move the lowest two (light-blue) simulation points in
Fig.~\ref{fig:IRgamma} closer to the calorimetric relation.

So far, we choose the canonical value for the CR injection efficiency at
supernovae of $\zeta_{\rmn{SN}}=0.1$. Varying this value by a factor of three
results in a similar change in $L_\gamma$ (Table~\ref{tab:Lgamma}).  In all
these cases, the CR injection and adiabatic gains are balanced by the total CR
cooling rate, indicating self-regulation. Hence, $L_\gamma$ and---by
extension---the FIR--gamma-ray relation remain invariant if $\zeta_{\rmn{SN}}$
decreases by a factor of two and the CR spectral index is increased to
$\alpha=2.15$ (Fig.~\ref{fig:gamma_spectrum}).

\section{Concluding remarks}
\label{sec:conclusion}

For the first time, we have calculated the gamma-ray emission in galaxy
simulations that span four orders of magnitude in SFRs and by modeling ideal MHD
and CR physics self-consistently. We identify the influence of different CR gain
and loss processes on the gamma-ray emission.

In agreement with previous literature, we find that the gamma-ray luminosity
from decaying pions dominates over the IC emission by at least a factor of 5.7
in the \Fermi-energy~band $0.1-100$~GeV. This dominance increases significantly
for spectral indices steeper than $\alpha=2.05$. The continuous injection of CR
energy at supernova remnants increases CR pressure to the point where the
buoyancy force overcomes the magnetic tension of the dominant toroidal field
after 1~Gyr. This enables CRs to diffuse into the halo and to accelerate the gas
in a bubble-like, CR-dominated outflow. However, these bubbles are invisible in
our simulated gamma-ray maps of hadronic pion-decay and inverse-Compton emission
because of low gas density in the outflows. This suggests that morphological
features such as the \Fermi bubbles in the Milky~Way may be generated by
leptonic IC emission.

We find that most CR energy is lost to hadronic interactions at high SFRs.
However, the gamma-ray emission deviates from this calorimetric property at low
SFRs due to adiabatic losses, which cannot be identified in traditional one-zone
models. Assuming that gamma rays mainly originate from decaying pions, we show
that our simulated galaxies exactly reproduce the observed FIR--gamma-ray
relation at FIR luminosities $L_{8-100~\mu\rmn{m}}>10^9~L_\odot$. This
non-trivial finding comes about because we adopt a phenomenological relation
between SFR and FIR emission and because
$L_{\pi^0-\gamma}\propto{f}_{\rmn{cal}}L_\CR\propto{f}_{\rmn{cal}}\rmn{SFR}$
where the calorimetric factor ${f}_{\rmn{cal}}$ smoothly decreases towards
smaller SFRs due to the increasing importance of adiabatic losses. At small FIR
luminosities ($L_{8-100~\mu\rmn{m}}<10^9~L_\odot$), the Magellanic~Clouds emit
less gamma rays in comparison to our simulated analogues. We speculate that this
is either due to a too dense ISM or due to a missing multi-phase, porous ISM in
our simulations, which would lower the hadronic gamma-ray yield at fixed gas
mass, provided the CR density is (anti-)correlated with the thermal gas.

Most importantly, the simulated gamma-ray emission does not depend on whether we
account for anisotropic CR diffusion in addition to CR advection. The somewhat
higher CR energy is compensated by the lower gas density in our {\em CR~diff}
model in comparison to the {\em CR~adv} model. This demonstrates that
uncertainties in modeling active CR transport processes only play a minor role
in predicting gamma-ray emission from galaxies and emphasizes the importance of
dynamic simulations of galaxy formation to understand non-thermal processes.

\begin{table*}
  \caption{SFRs and gamma-ray luminosity $L_{0.1-100\,\rmn{GeV}}$ for our simulated galaxies}
  \begin{center}
\begin{tabular}{ccc | ccc | ccc}
\hline
\hline
 & & & \multicolumn{3}{c|}{{\em CR~diff} model} & \multicolumn{3}{c}{{\em CR~adv} model}\\ 
 $M_{200}$  & $c_{200}$& $\zeta_{\rmn{SN}}$ & $t$ & SFR & $L_{\pi^0-\gamma}$ & $t$ & SFR & $L_{\pi^0-\gamma}$\\
\phantom{\big|}%
[$\msun$] & & & [Gyr] & [$\msun~\rmn{yr}^{-1}$] & [$\rmn{erg~s}^{-1}$] & [Gyr] & [$\msun~\rmn{yr}^{-1}$] & [$\rmn{erg~s}^{-1}$] \\
\hline\\[-1em]
 $10^{12}$  & 12  & 0.10 & 0.1 & $1.2\times10^2$~~ & $2.76\times10^{41}$ & ~~0.1 & $9.3\times10^1$~~  & $1.85\times10^{41}$ \\
 $10^{12}$  & 12  & 0.10 & 0.3 & $4.5\times10^1$~~ & $1.01\times10^{41}$ & ~~0.4 & $3.8\times10^1$~~  & $8.43\times10^{40}$ \\
 $10^{12}$  & 12  & 0.10 & 1.0 & $1.7\times10^1$~~ & $3.91\times10^{40}$ & ~~1.2 & $1.3\times10^1$~~  & $2.72\times10^{40}$ \\
 $10^{12}$  & 12  & 0.10 & 2.3 & $6.0\times10^0$~~ & $1.31\times10^{40}$ & ~~2.5 & $4.5\times10^0$~~  & $9.08\times10^{39}$ \\
 $10^{12}$  & 12  & 0.10 & 4.8 & $2.1\times10^0$~~ & $5.19\times10^{39}$ & ~~4.5 & $1.7\times10^0$~~  & $3.54\times10^{39}$ \\
 $10^{11}$  & 12  & 0.10 & 0.2 & $6.3\times10^0$~~ & $1.45\times10^{40}$ & ~~0.4 & $5.3\times10^0$~~  & $9.76\times10^{39}$ \\
 $10^{11}$  & 12  & 0.10 & 0.7 & $2.4\times10^0$~~ & $4.04\times10^{39}$ & ~~0.9 & $1.8\times10^0$~~  & $3.36\times10^{39}$ \\
 $10^{11}$  & 12  & 0.10 & 1.3 & $7.4\times10^{-1}$ & $1.08\times10^{39}$ & ~~1.8 & $7.2\times10^{-1}$ & $1.26\times10^{39}$ \\
 $10^{11}$  & 12  & 0.10 & 2.2 & $3.0\times10^{-1}$ & $4.08\times10^{38}$ & ~~3.0 & $3.2\times10^{-1}$ & $5.41\times10^{38}$ \\
 $10^{10}$  & 12  & 0.10 & 0.1 & $1.6\times10^{-1}$ & $1.31\times10^{38}$ & ~~0.2 & $1.9\times10^{-1}$ & $1.96\times10^{38}$ \\
 $10^{10}$  & 12  & 0.10 & 0.6 & $6.1\times10^{-2}$ & $6.13\times10^{37}$ & ~~2.0 & $7.1\times10^{-2}$ & $1.07\times10^{38}$ \\
 $10^{10}$  & 12  & 0.10 & 1.2 & $2.1\times10^{-2}$ & $1.63\times10^{37}$ &  12.0 & $3.0\times10^{-2}$ & $5.26\times10^{37}$ \\
 $10^{10}$  & 12  & 0.10 & 2.0 & $7.6\times10^{-3}$ & $4.67\times10^{36}$ & & & \\ 
 $10^{12}$  & ~~7 & 0.03 & 0.2 & $5.7\times10^1$~~ & $3.93\times10^{40}$ & & & \\
 $10^{12}$  & ~~7 & 0.03 & 0.8 & $2.2\times10^1$~~ & $1.78\times10^{40}$ & & & \\
 $10^{12}$  & ~~7 & 0.03 & 2.3 & $7.8\times10^0$~~ & $6.31\times10^{39}$ & & & \\
 $10^{12}$  & ~~7 & 0.05 & 0.2 & $5.7\times10^1$~~ & $6.36\times10^{40}$ & & & \\
 $10^{12}$  & ~~7 & 0.05 & 0.7 & $2.2\times10^1$~~ & $2.80\times10^{40}$ & & & \\
 $10^{12}$  & ~~7 & 0.05 & 2.1 & $7.7\times10^0$~~ & $8.64\times10^{39}$ & & & \\
 $10^{12}$  & ~~7 & 0.10 & 0.2 & $5.8\times10^1$~~ & $1.33\times10^{41}$ & & & \\
 $10^{12}$  & ~~7 & 0.10 & 0.6 & $2.2\times10^1$~~ & $5.01\times10^{40}$ & & & \\
 $10^{12}$  & ~~7 & 0.10 & 1.9 & $7.6\times10^0$~~ & $1.54\times10^{40}$ & & & \\
 $10^{12}$  & ~~7 & 0.30 & 0.2 & $5.8\times10^1$~~ & $3.31\times10^{41}$ & & & \\
 $10^{12}$  & ~~7 & 0.30 & 0.5 & $2.2\times10^1$~~ & $1.22\times10^{41}$ & & & \\
 $10^{12}$  & ~~7 & 0.30 & 1.6 & $7.9\times10^0$~~ & $4.72\times10^{40}$ & & & \\
\hline
\end{tabular}
\end{center}
\label{tab:Lgamma}
\end{table*}

\section*{Acknowledgements}

We thank Else Starkenburg, Annette Ferguson and Alice Quillen for useful
discussions and an anonymous referee for a constructive report. CP, RP, CS, and
VS acknowledge support by the European Research Council under ERC-CoG grant
CRAGSMAN-646955, ERC-StG grant EXAGAL 308037 and from the Klaus Tschira
Foundation.

\bibliographystyle{aasjournal} 

\label{lastpage}
\end{document}